\documentclass[preprint,prb]{revtex4}
\usepackage{graphicx}
\usepackage{color}
\begin{document}
\bibliographystyle{unsrt}
\title{Growth Velocities of Branched Actin Networks}
\author{A. E. Carlsson}
\affiliation{Department of Physics, Washington University,
St. Louis, Missouri 63130-4899}
\begin{abstract}
The growth of an actin network against an obstacle
that stimulates branching locally
is studied using several variants of a kinetic rate
model based on the orientation-dependent number density of
filaments.
The model emphasizes the effects of branching and
capping on the density of free filament ends.
The variants differ in their treatment of
side vs. end branching and dimensionality, and
assume that new branches are generated
by existing branches (autocatalytic behavior) or
independently of existing branches (nucleation behavior).
In autocatalytic models, the network growth velocity is 
rigorously independent of the opposing 
force exerted by the obstacle, and the network density 
is proportional to the force. The dependence of the growth
velocity on the branching and capping rates is evaluated
by a numerical solution of the rate equations. 
In side-branching models,
the growth velocity drops gradually to zero with decreasing
branching rate, while in end-branching models the drop is
abrupt. As the capping rate goes to zero, it is found that
the behavior of the velocity is sensitive to
the thickness of the branching region. Experiments are
proposed for using these results to shed light on the
nature of the branching process. 
\end{abstract}
\maketitle

\newpage

\section{Introduction}

In numerous instances of actin-based motility, including extension of
lamellipodia in cells\cite{Small02}, and ``rocketing" motion of {\it Listeria
monocytogenes}\cite{Dramsi98,Goldberg01} or small beads coated with 
actin-polymerization activators\cite{Cameron01,Bernheim02}, 
actin filaments form a branched network structure. Typical
densities of filamentous actin in such networks are 1 mM,
and spacings between branches along a filament are 
often in the range 40-70 nm\cite{Svitkina97,Svitkina99}.
The filaments are eventually terminated by capping proteins, which
prevent further filament growth.
The branch points have a characteristic angle of
$70^{\circ}$ and are decorated by a seven-subunit complex
of actin-related proteins, Arp2/3.
This complex has a low constitutive activity. 
However, it can be activated directly or
indirectly by several agents associated with the obstacle (the cell
membrane or the bacterial/bead surface). These agents include the 
membrane phospholipid $PIP_2$, the membrane-associated protein $Cdc42$,
and the bacterial surface protein $ActA$. 
In the case of $Cdc42$ and $PIP_2$, intermediate proteins such as Scar 
and WASp are required for Arp2/3 activation. These are also constitutively 
inactive but can be activated by $Cdc42$ or $PIP_2$.
When the Arp2/3 is activated, it causes new branches to form 
on existing filaments and thereby greatly stimulates
actin polymerization in the vicinity of the obstacle. 
The branching activity of Arp2/3 has been confirmed by
{\it in vitro} studies\cite{Mullins98}. 

While several
of the basic biochemical events in the pathway leading to 
Arp2/3-induced actin assembly are well established, the details of the 
process by which new filaments are generated at the obstacle are 
not well understood. The following issues are among those that
are unresolved:
\begin{itemize}
\item{The relative importance of branching along filament sides
and branching at their ends. 
Initial data\cite{Pantaloni00} comparing the lengths of 
mother and daughter filaments (beyond the branch point) found
a close correlation, suggesting the dominance of end branching. 
However, recent total internal-reflection fluorescence 
microscopy studies\cite{Amann01,Amann01a} of single filaments
have found that most branches are formed along filament sides. 
A recent confocal-microscopy study\cite{Ichetovkin02} found
that branches can form anywhere along the sides of filaments,
but that there were numerous instances of new branches forming
very near the barbed end, and branches formed more readily
on newly grown filaments. 
This suggested that branch formation could be 
enhanced in the ``ATP cap" region near a filament barbed end.} 
\item{The thickness of the region near the obstacle where
new branches can form. Branch formation could, for example,
be activated by direct contact with membrane proteins; on 
the other hand, Arp2/3 could be activated by membrane-bound proteins
and subsequently diffuse to the branching point, or it could
be indirectly activated by effectors of these proteins. 
In a recently proposed  model for filament generation at 
membranes\cite{Wear00}, 
$Cdc42$ and $PIP_2$ in the membrane are first activated by
an external signal. They interact with WASp, causing
it to change to a partly active conformation. Binding of actin
to WASp completes its activation.
Then the WASp binds to, and activates, the
Arp2/3 complex, which is also associated with a filament. 
Finally, a new filament grows from the activated Arp2/3 complex. 
In this model, if the WASp is attached to the 
$Cdc42$ and $PIP_2$, branching could only occur if
the Arp2/3 is essentially in physical contact with the
obstacle; if the WASp detaches, the branching region 
could be wider.}
\item{Whether new filaments are created on existing filaments,
or are created free and subsequently diffuse and attach to 
existing filaments (cf. Figure 1). We term these models the
autocatalytic and nucleation models, respectively. In autocatalytic
models, the formation rate of new branches is proportional to 
the number of filaments or amount of polymerized actin in the
branching region; in nucleation models, it is independent of
the number of preexisting branches. The scenario\cite{Wear00} 
discussed above leads to an autocatalytic behavior if the concentrations 
of Arp2/3 and WASp are not rate-limiting, since the 
Arp2/3 is filament-asociated.  
Such autocatalytic models have generally been in favor because
of the enhancement of Arp2/3 {\it in vitro} nucleating activity in
the presence of preformed filamentous actin\cite{Machesky99a,Higgs99,
Pantaloni00}. In the absence of preformed filaments, actin 
polymerization in the presence of activated Arp2/3 typically 
has a lag time on the order of minutes; this lag time is eliminated
by the presence of preformed filaments. We also note that 
{\it in vitro} polymerization kinetics are well described by
autocatalytic models\cite{Pantaloni00}. It is plausible that 
the generation of new filaments in lamellipodia and around intracellular
pathogens/beads involves essentially the same steps as the 
{\it in vitro} studies. However, the sequence of steps may not
necessarily be the same as in the {\it in vitro} studies. 
If the concentration of Arp2/3 or its activators is rate-limiting,
then the generation rate for new filaments will be 
nearly independent of the filament concentration. 
At present, there appears to be no
straightforward experimental method for distinguishing between
the autocatalytic and nucleation models {\it in vivo}. 
The true behavior is very likely 
somewhere between the limiting cases defined here, but these
cases form a useful conceptual framework. }
\end{itemize}

The main purpose of this paper is to evaluate the dependence
of the growth velocity of actin networks on key protein concentrations
and opposing force, and to ascertain
how these dependences are modulated by key
molecular-scale details of the branching process, including
the relative importance of side and end branching,
the thickness of the branching region, and whether the branching
process is autocatalytic or nucleation-driven. 
These predicted dependences can be 
combined with experimental measurements to establish important 
molecular aspects of the branching process. 
Because the autocatalytic model is more plausible in view
of existing experimental data, we treat it in more detail;
our treatment of the nucleation model is mainly focused
on distinguishing it from the autocatalytic model experimentally. 
We have previously evaluated\cite{Carlsson01} the dependence of
the growth velocity on branching rate,
capping rate, and opposing force, 
for an autocatalytic branch generation model, 
using a stochastic simulation methodology. These simulations showed 
that over a limited range of parameters at fixed actin
concentration, the growth velocity
drops linearly with increasing capping rate, 
and drops to zero for values of
the branching rate at which the number of 
branches per filament is less than about 1.5. 
It was also found that
the growth velocity is nearly independent of the applied force.
The present paper treats some of the same issues using a
deterministic rate-equation model. This approach has three
advantages over the stochastic-growth approach. First, it 
is possible to prove rigorous results within such a model, 
and this explains some puzzling results of the
stochastic-growth simulations. Second, it is possible to treat
parameter regimes that were computationally forbidding using
the stochastic-simulation methodology, in particular the limits
of small capping-protein concentration and slow growth. The
former leads to an unwieldy number of branches 
per filament, and the latter results are very sensitive to stochastic
fluctuations in the simulations. Finally, by 
analysis of the deterministic equations, it is possible to find 
intuitive explanations for the behavior of the growth velocity. 

\section{Autocatalytic Model}

Our model assumes a flat
obstacle of finite size in two or three dimensions, although the
rigorous results that we prove hold for an obstacle of arbitrary shape. 
The branching mechanism is such that 
new filaments are generated inside a narrow branching region,
of thickness $d$, at the obstacle.  
Only filaments within a distance $d$ of the obstacle can branch.
In a rigorously two-dimensional model, $d$ would be a width, but
since there is a always a third dimension present, we will still
call $d$ a thickness.
The mathematical approach uses simple rate 
equations based on the laterally averaged filament orientation
distribution ${\rm n(\theta ,t)}$, where $t$ is time, and 
$\theta$ is the angle of a filament with respect to the normal
to the surface (cf. Figure 1).  
The component $v$ of the filament 
growth velocity in the direction of network
growth is related to the orientation
by $v(\theta )= V_{\rm max}\cos{\theta}$, where 
$V_{\rm max}$ is the growth velocity of a free filament. 
The number of filament ends in 
the branching region, per unit of obstacle length (in two dimensions)
or obstacle area (in three dimensions) with angles 
between $\theta$ and $\theta + d\theta$ is $n(\theta,t)d\theta$.
The main factors of interest to us are the formation of new filaments
inside the branching region, the capping of existing filaments,
and the motion of the obstacle away from the filaments in
the branching region. We ignore potential effects of uncapping
and branch detachment; the rationale for this is discussed in
the section {\bf Sensitivity to Key Approximations}.

We thus employ the following equation of motion for the filament
orientation distribution:
\begin{eqnarray}
 {\partial n(\theta,t) \over \partial t} &=& k_{\rm br} 
\int_{0}^{\theta_{\rm max}}
D(\theta ,\theta ') \nu(\theta') n(\theta ',t) d\theta '
-k_{\rm cap}n(\theta ,t) 
\nonumber \\ 
 & & - H[V_{\rm obst}-v(\theta)]
[(V_{\rm obst}-v(\theta))/d +(\nu(\theta)k_{\rm br}-k_{\rm cap})]n(\theta,t),
\label{dndt}
\end{eqnarray}
Here $\theta_{\rm max}$ is the 
maximum value of $\theta$ (taken to be 
$180^{\circ}$ for most of our calculations), 
$k_{\rm br}$ is the total branching rate for a filament
with $\theta=0$, 
$k_{\rm cap}$ is the capping rate, $D(\theta ,\theta ')$ is the
distribution of filament orientations 
generated by branching from filaments of orientation 
$\theta '$, and $\nu(\theta)$ is a factor describing the
dependence of filament length on orientation in 
side-branching models; in end-branching models,
we take $\nu(\theta)=1$. 
$V_{\rm obst}$ is the obstacle velocity, and
$\rm H$ is the Heaviside step function, defined by 
$\rm H [V_{\rm obst}-v(\theta)] = 1$ if 
$V_{\rm obst}>v(\theta)$ and  
$\rm H [V_{\rm obst}-v(\theta)] = 0$ if 
$V_{\rm obst}\le v(\theta)$.
For end branching models, we assume that, unless 
$\theta$ is restricted, the overall rate of
branching from a given filament is independent of $\theta'$.
Thus $\int_{0}^{180^{\circ}} D(\theta ,\theta ') d\theta =1$ 
for all $\theta '$.  For side-branching models, we assume that
the rate of branching from a given filament is proportional
to the length of its portion inside the branching region,
as described by the $\nu (\theta)$ term. Detailed forms for
$D(\theta,\theta')$ and $\nu (\theta)$ are given in the next section. 
The last term on the right-hand side describes filaments 
with $v < V_{\rm obst}$ leaving the branching
region. The rate of this process is proportional to the spatial 
number density
of filament ends at the back end of the branching region. 
(Here and in the rest of the paper, the term ``density" will
always refer to number density rather than mass density.) For
most values of $\theta$, the relative velocity of the filaments and
the obstacle is large enough that the distribution is
fairly constant in space; in this case the density can be  
approximated by $n(\theta ,t)/d$, leading to the first term in
square brackets.  However, filaments with $v$ very
close to $V_{\rm obst}$ can remain in the branching region long enough
that the density at back of the branching region greatly exceeds
that at the front, because of exponential growth due to
branching (and modified by capping). For such filaments, 
we assume a time growth rate of 
$k_{\rm br}\nu (\theta ) -k_{\rm cap}$, 
leading to a spatial growth rate of
($k_{\rm br}\nu (\theta ) -k_{\rm cap})/(V_{\rm obst}-v)$. 
This yields the second term inside the brackets. 
For filaments with
$v > V_{\rm obst}$, we include no leaving terms. 
Filaments are not
able to leave at the front end of the branching region because
they are blocked by the obstacle, and they cannot grow in from
the back end, since branching cuts off there so that no
new filaments can be nucleated beyond that point.

This model is closely related to one previously 
employed\cite{Maly01} in the calculation of actin filament 
orientation distributions near obstacles. The main difference
is that effects of filaments leaving the branching region 
are treated explicitly in the present
model. This allows us to study the mechanism for establishing the 
steady-state number of filaments and velocity; in Ref. \onlinecite{Maly01},
these were treated as fixed inputs. Our model is also related
to those studied in Ref. \onlinecite{Mogilner03a} and 
Ref. \onlinecite{Mogilner02b}; the parallel is 
explored in more detail in the section describing nucleation models. 

\bigskip
\subsection{Rigorous Properties of Rate Equation}

In this section, we demonstrate two rigorous steady-state properties 
of the rate equation (\ref{dndt}): that the growth velocity
is independent of the applied force, and that the network
density is proportional to the applied force. These results hold
regardless of the form of the branching orientation distribution
$D(\theta,\theta')$, and are independent of the shape of the
obstacle. Before proving the results, 
we first clarify the mechanism by which 
the steady-state $V_{\rm obst}$ is determined. Figure 2
sketches the generic behavior. We consider the limits 
$V_{\rm obst}=0$ and $V_{\rm obst}=V_{\rm max}$ first.
If $V_{\rm obst} =0 $, then the
leaving terms vanish  for filaments with
$\theta < 90^{\circ}$. If we define $k_{\rm br}^{*}$ 
as the rate of branching restricted to the subset
of filaments with $\theta < 90^{\circ}$, 
then Eq. (\ref{dndt}) implies that
the total number of filaments touching the obstacle,
$n_{\rm touch}(t) = \int_{0}^{90^{\circ}} 
n(\theta ,t) d\theta$, satisfies
\begin{equation}
dn_{\rm touch}/dt \ge (k_{\rm br}^{*}-k_{\rm cap})n_{\rm touch}(t),
\label{dntotdt}
\end{equation}
where the relation is an inequality because additional touching
branches can be produced by non-touching filaments. 
Eq. (\ref{dntotdt}) gives exponential growth 
if $k_{\rm br}^{*}> k_{\rm cap}$. 
This inequality holds for the networks that
have been studied by electron microscopy. Since the
$\theta > 90^{\circ}$ region includes half of the angles available
for branching, $k_{\rm br}^{*}$ is roughly half of $k_{\rm br}$. 
The ratio $k_{\rm br}/k_{\rm cap}$, in turn, is the ratio of the filament
length to the branch spacing, and this is five or more in the
observed structures\cite{Svitkina97,Svitkina99}. Thus
$dn_{\rm touch}/dt > 0$ for $V_{\rm obst} = 0$.
On the other hand, when $V_{\rm obst}$ is very 
close to  $V_{\rm max}$, only a very small fraction of newly
generated filaments will touch the obstacle, so capping
and leaving terms will dominate.
Then $dn_{\rm touch}/dt < 0$. 
The value of $dn_{\rm touch}/dt$
will then cross zero at a value of $V_{\rm obst}$ between 
$0$ and $V_{\rm max}$, and this determines the 
steady-state velocity. 

The above discussion is somewhat incomplete because 
$dn_{\rm touch}/dt$ is determined not just by $n_{\rm touch}$,
but by the entire distribution $n(\theta ,t)$. The time evolution
of $n(\theta ,t)$ can be described more precisely by noting that
the growth or decay of the filament density is determined
by the eigenvalues of the right-hand side of Eq. (\ref{dndt}). 
The largest eigenvalue will dominate at large times. For 
$V_{\rm obst} = V_{\rm max}$, all of the eigenvalues are
negative and the solution decays. For $V_{\rm obst} = 0$,
there will be a positive eigenvalue, and the filament density
will grow exponentially. At a critical value of $V_{\rm obst}$,
the largest eigenvalue will cross zero, and this is the steady-state
value of $V_{\rm obst}$. The filament orientation distribution is 
proportional to the eigenvector corresponding to the zero eigenvalue, 
as was noted by Ref. \onlinecite{Maly01}. 

To show that the growth velocity is independent of the
applied force, we first demonstrate that
if $n(\theta )$ is a steady-state solution, then any multiple of 
$n(\theta )$ is also a steady-state solution. This follows immediately
from the form of Eq. (\ref{dndt}), since each term is linear
in $n(\theta ,t)$. 
We then write the total force exerted by the filaments on the obstacle as 
\begin{equation}
F_{\rm obst} = \int_{0}^{\theta_{\rm obst}} 
f(v(\theta) ;V_{\rm obst}) n(\theta ) d\theta
\times ({\rm ~ area ~ or ~ length}),
\label{ftot}
\end{equation}
where $f(v(\theta);V_{\rm obst})$ is the force exerted by a filament
at an angle $\theta$ on an obstacle moving at velocity 
$V_{\rm obst}$, $\theta_{\rm obst}$ is defined by
$v(\theta_{\rm obst}) = V_{\rm obst}$,
and the total force contains a factor of either
area or length according to whether the model is three- or
two-dimensional. Filaments growing at angles greater than 
$\theta_{\rm obst}$ exert no force since they do not remain in 
contact with the obstacle. 
If a given set $[n(\theta),F,V_{\rm obst}]$ gives a steady-state
solution, then Eqs. (\ref{dndt}) and (\ref{ftot}) show 
that for any $\alpha$ the set $[\alpha n(\theta ),\alpha F,V_{\rm obst}]$ will
also give a steady-state solution. Thus for any force 
$\alpha F$, the steady-state velocity will be $V_{\rm obst}$; 
the filament orientation distribution is $\alpha n(\theta)$,
and is thus proportional to the applied force. 

The physical scenario leading to the
obstacle velocity being independent of the applied force is that
when the force on the obstacle is increased, the obstacle will temporarily
slow, allowing the creation of new filaments, until the filament density
is precisely that required to compensate for the additional obstacle
force. Then the velocity returns to its steady-state value. 
The present results explains the corresponding results
found in the stochastic-growth simulations Ref. \onlinecite{Carlsson01}, 
which were not previously understood.
The result obtained here is significantly more general than that 
obtained in these simulations.  The only aspect of Eq. (\ref{dndt}) used to 
derive the result is that all the terms are linear in ${\rm n}(\theta )$.
Therefore, the result would also hold for obstacles of arbitrary
shape, for which  $D(\theta,\theta')$ would also depend on the position
of the branching filament. 
It also holds for any form of the interaction force between the filaments
and the obstacle. It continues to hold when several effects not included in
the present model are included, but its validity will be limited when
filament-filament interactions, depletion of actin and actin-binding
proteins, and filament-number fluctuation effects
are important.  These aspects of the results are discussed 
in the section {\bf Sensitivity to Key Approximations}.

\bigskip
\newpage
\subsection{Numerical Solution of Rate Equation}

Evaluation of the dependence of the growth velocity on the rate 
parameters $k_{\rm br}$ and $k_{\rm cap}$ provides several
avenues for comparing the model predictions with experimental data. 
These parameters should correspond roughly to the concentration
of activated $Arp2/3$ complex and capping protein. However, 
the correspondence is not exact, since changes in the concentrations 
of these proteins
can lead to changes in the free actin monomer concentration
and thus change $V_{\rm max}$, 
as well as $k_{\rm br}$.  In addition, the net branching and
capping rates will be determined by capping-uncapping
and branching-debranching equilibria, which do not
give a strictly linear dependence of the rates on
the protein concentrations. 
In order to evaluate the dependence of the growth velocity
on $k_{\rm br}$ and $k_{\rm cap}$, we solve
the rate equation numerically, using four different forms 
for $D(\theta ,\theta ')$: 

\begin{itemize}
\item{{\bf Two-dimensional geometry, end branching.} In this
geometry, we assume a Gaussian spread of the branching angle
of width $\rm \Delta \theta = 10^{\circ}$ with respect to
its average value $\theta_{\rm br} = 70^{\circ}$. This value
is a rough mean of the measured widths of the distribution in
{\it in vitro} experiments, which range from
$7^{\circ}$\cite{Mullins98} to 
$10^{\circ}-13^{\circ}$\cite{Blanchoin00a}. Then
\begin{eqnarray}
&&D(\theta ,\theta ')  =
[\exp(-(\theta-\theta'-\theta_{\rm br})^2/2\Delta \theta^2)+
\exp(-(\theta-\theta'+\theta_{\rm br})^2/2\Delta \theta^2)+ \nonumber \\
&& \exp(-(\theta+\theta'-\theta_{\rm br})^2/2\Delta \theta^2)+
\exp(-(\theta+\theta'+\theta_{\rm br})^2/2\Delta \theta^2)]/
(32 \pi)^{1/2} \Delta \theta.
\label{2d}
\end{eqnarray}
The alternating plus and minus signs
preceding $\theta_{\rm br}$ correspond to branching in clockwise and
counterclockwise directions, and those in front of $\rm \theta'$ 
account for branching from the right semicircle to the left
semicircle.}

\item{{\bf Three-dimensional geometry, end branching.} In this
geometry, we assume the same values of 
$\theta_{\rm br} = 70^{\circ}$
and $\rm \Delta \theta = 10^{\circ}$ as in the two-dimensional
model. However, in three dimensions the orientation distribution is
more complicated because different values of the azimuthal angle
$\phi$ (the angle describing rotation about the mother filament) 
of the new filament give different values of $\theta$. One readily
shows that 
\begin{eqnarray}
\cos{\theta} = \cos{\theta_{\rm br}}\cos{\theta'}+
\sin{\theta_{\rm br}}\cos{\phi}\sin{\theta'}, 
\label{phi}
\end{eqnarray}
where $\phi = 0$ is chosen to be in the plane defined by
the orientations of the mother filament and the normal to
the obstacle. Then 
\begin{equation}
D(\theta,\theta')=(\sin{\theta}/\pi)
\int_0^{\pi} \delta [\cos{\theta} -
\cos{\theta_{\rm br}}\cos{\theta'}-
\sin{\theta_{\rm br}}\cos{\phi}\sin{\theta'}] d\phi,
\end{equation}
which, after simplification, yields 
\begin{equation}
D(\theta,\theta')=\sin{\theta}/[\pi 
\sin{\theta_{\rm br}} \sin{\theta'} \sin{\phi}],
\label{3d}
\end{equation}
where $\phi$ is determined by Eq. (\ref{phi}) and
we choose $0 \le \phi \le 180^{\circ}$.
We include the broadening by writing 
$D(\theta,\theta')$ as a linear combination of terms of the form
given by Eq. (\ref{3d}), for closely spaced set of values of
$\theta_{\rm br}$, with weights determined by the Gaussian 
distribution used in Eq. (\ref{2d}).}
\item{{\bf Two-dimensional geometry, side branching.} When
side branching is present, 
filaments nearly parallel to the obstacle will have a
greater length inside the branching region, and will thus
branch more rapidly than those perpendicular to the 
obstacle. We take this into account by multiplying the 
end-branching result of Eq. (\ref{2d}) by the following 
angular factor, which takes different forms for filaments
with $v>V_{\rm obst}$ and those with $v<V_{\rm obst}$:
\begin{eqnarray}
\nu(\theta') =&{\rm min}[1/\cos{\theta'},V_{\rm max}/k_{\rm cap}d] 
~~ & v>V_{\rm obst} \nonumber \\
\nu(\theta') =&{\rm min}[1/\cos{\theta'},V_{\rm max}/k_{\rm cap}d,
V_{\rm max}/4(V_{\rm obst}-v(\theta'))] ~~  & v<V_{\rm obst}
\label{side}
\end{eqnarray}
This factor is approximately equal to the average filament
length, divided by $d$. 
In the first case, the filament ends are in contact with the
obstacle, and the $1/\cos{\theta}$ term comes from the length of
the piece of an infinitely long filament that is inside the
branching region; the $V_{\rm max}/k_{\rm cap}d$ term accounts
for the finiteness of the filament length induced by capping.
In the second case, the last term accounts for the fact that
the filament length is limited by the amount of time it has
spent in the branching region. A calculation assuming uniform
filament distributions shows that the average age of a
filament in the branching region is $d/4(V_{\rm obst}-v(\theta'))$.}
\item{{\bf Three-dimensional geometry, side branching.} The
three-dimensional end-branching result, Eq. (\ref{3d}), is
multiplied by the same factor $\nu(\theta')$ as in the two-dimensional
case.} 
\end{itemize}

To obtain the obstacle velocity, the integral in Eq. (\ref{dndt})
is replaced by a Riemann sum over a finely spaced set of values of
$\theta$ and $\theta'$. This converts it into a matrix equation 
of the form $dn_i/dt=\sum_j A_{\rm ij}n_j$, where the coefficients
$A_{\rm ij}$ include all of the terms in Eq. ({\ref{dndt}).
A standard eigenvalue finder 
(``dgees.f" in the Lapack library\cite{Lapack99})
is used to find the eigenvalues of the matrix $A_{\rm ij}$. 
They are monotonically decreasing as a function of $V_{\rm obst}$,
and a search is made over a range of possible values of $V_{\rm obst}$
to find the value of $V_{\rm obst}$ at which the largest eigenvalue
is closest to zero, which gives $dn_i/dt=0$ and thus leads to 
steady-state behavior. 

The results are plotted in Figure 3 vs. $k_{\rm br}$ and Figure 4
vs. $k_{\rm cap}$. In generating Figure 3, we use a fixed value of
$0.35 sec^{-1}$ for $k_{\rm cap}$. This is obtained
from measured {\it in vitro} capping
rate constants\cite{Schafer96} of about $3.5 \mu M^{-1} sec^{-1}$
and a typical capping-protein concentration\cite{Pollard00}
of $1 \mu M$, on the assumption that diffusion in the cellular
environment is slower than that {\it in vitro} by a factor
of 10, with a corresponding reduction in the capping rate.  
Our value of $V_{\rm max}$, $0.27 \mu m~sec^{-1}$, is obtained from the
measured on-rate\cite{Pollard86} of roughly 
$10 \mu M^{-1} sec^{-1}$, a typical free-actin 
concentration\cite{Pollard00} of $100 \mu M$, and the
monomer step size of $2.7~nm$, together 
with the diffusion-factor reduction used in obtaining $k_{\rm cap}$. 
(Most of the non-polymerized actin in cellular environments is present
as profilin-actin complexes, and it is not known at what rate
actin in this form contributes to filament elongation. If its
addition rate is much less than that for actin monomers, 
then the overall scale of the velocities will be reduced. 
However, when $V_{\rm obst}$ is scaled by $V_{\rm max}$ as
in Figures 3 and 4, the shape of the curves is independent of $V_{\rm max}$.)
We normalize $k_{\rm br}$ by $k_{\rm cap}$ because the ratio corresponds 
roughly to the average number of branches per filament. 
In these curves and those in Figure 4, we use $d=a$, where
$a=2.7nm$ is the step size per monomer; the effects of
varying $d$ are treated in the discussion of Figure
5 below.
In both the 
two- and three-dimensional 
cases, $V_{\rm obst}$ appears to approach an asymptotic value less 
than $V_{\rm max}$ for large $k_{\rm br}$, as was seen
in the stochastic-growth simulations\cite{Carlsson01}. 
For small $k_{\rm br}$,
the velocity in the end-branching case drops abruptly to zero at a value
of $k_{\rm br}$ between $k_{\rm cap}$ and $2k_{\rm cap}$;
for side branching, the decrease is smoother. 
This difference was not resolved in the stochastic-growth 
simulations because they did not treat long enough times. 
In the two-dimensional geometry, the curve 
has a shoulder around 
$k_{\rm br}$ = $2k_{\rm cap}$, which is not seen in
the three-dimensional results. We believe that
this shoulder is due to the presence of sharp peaks in the
filament orientation distribution around $\pm 35^{\circ}$. 
Such peaks were seen in the analysis of Ref. \onlinecite{Maly01}, 
and we see similar peaks here. 

In Figure 4, we use 
$k_{\rm br} = V_{\rm max}/20a = 5 {\rm sec}^{-1}$, which 
gives a branch spacing of about 20 monomers at the higher obstacle 
velocities, roughly commensurate with experimentally measured
branch spacings\cite{Svitkina99}. For all four of the branching
models considered, $V_{\rm obst}$ drops monotonically
and smoothly with $k_{\rm cap}$ for 
$k_{\rm cap} > 0.05  k_{\rm br}$, as in the stochastic-growth
simulations.  The asymptotic 
$k_{\rm cap} \rightarrow 0$ value extrapolated from this range
is between $0.8V_{\rm max}$ and $0.9V_{\rm max}$. However, 
for smaller values of $k_{\rm cap}$, the curve turns up, and
approaches $V_{\rm max}$ as $k_{\rm cap} \rightarrow 0$; this
effect was not seen in the stochastic-growth simulations because
such small values of $k_{\rm cap}$ could not be treated. 

We can understand these aspects of the behavior of the
growth velocity as follows:
\begin{itemize}
\item{Vanishing of growth velocity at finite 
$k_{\rm br}/k_{\rm cap}$. We note that the only positive
term in Eq. (\ref{dndt}) is the branching term. Therefore,
when $k_{\rm br}=k_{\rm cap}$ in the end-branching models, 
the total number of filaments must decay because  
leaving terms make a negative contribution to its time 
derivative (we recall that 
$\int_{0}^{180^{\circ}}
D(\theta ,\theta ') d\theta =1$). 
Thus no growth is possible
for $k_{\rm br}/k_{\rm cap} \le 1$. The actual threshold is
greater, because of the leaving effects. For side-branching
models, the situation is different because of the $\nu(\theta')$
factor in Eq. (\ref{dndt}). This factor can be significantly
greater than unity for filaments with $\theta$ near $90^{\circ}$,
which means that in principle growth is possible for
$k_{\rm br}/k_{\rm cap} \le 1$. As the obstacle slows, the
proportion of filaments with $\theta$ near $90^{\circ}$
increases because they can remain in contact with the obstacle,
and this causes the branching rate per filament to increase. 
This explains the small-$k_{\rm br}$ tail 
in the side-branching results.} 
\item{{\bf Asymptotic velocity.} One would expect that as
either the branching rate becomes infinite, or the capping rate
becomes small, sufficiently many filaments would be generated
that even the small fraction of the filaments with $\theta \simeq 0$
would be able to push the obstacle, leaving the other filaments behind. 
This would give an asymptotic velocity of $V_{\rm max}$.
This is seen in Figure 4 for very small values 
of $k_{\rm cap}$, but the
apparent asymptotic velocity extrapolated from higher values is less
than $V_{\rm max}$. To understand this crossover behavior, 
we note that the branching events can be divided into two types,
those occurring on filaments touching the obstacle, having 
$v> V_{\rm obst}$, and those occurring on filaments in the
branching region but not touching the obstacle, which have
$v< V_{\rm obst}$. As above, we will denote the number of 
filaments touching the obstacle by $n_{\rm touch}$. In steady state,
contributions to $n_{\rm touch}$ from branching are cancelled
by capping effects alone, since the leaving terms do not apply to
the touching filaments.  The branching contribution consists of
``direct" branching events in which a touching filament is
generated from another touching filament, and ``indirect" events
in which a touching filament is generated from a non-touching
one. In general, we expect direct events to dominate if
$d$ is much less than the typical branch spacing, since most
newly generated non-touching filaments will not have time
to branch before they leave the branching region. If we
ignore indirect events entirely, and take $\Delta \theta =0$
for simplicity, then the angle
$\theta$ of some of the touching filaments must exceed 
$\theta_{\rm br}/2=35^{\circ}$ for direct branching events to occur. 
This means that
$V_{\rm obst} \le \cos{(35^{\circ})}V_{\rm max} = 0.87 V_{\rm max}$,
giving an asymptotic value less than $V_{\rm max}$. This
explains the main parts of the curves in Figures 3 and 4. 

However, if $k_{\rm cap}$ is very small, it is possible for
the indirect events to dominate. Even though they are a small
fraction of the total branching events, they can be sufficient
to cancel a small $k_{\rm cap}$. For indirect events, there is
no geometrical limit on the obstacle velocity. For example, a filament
with $\theta =0$ will produce a daughter filament with
$\theta = 70^{\circ}$, and this filament can produce its own
daughter filament with $\theta = 0$. Thus for very small
values of $k_{\rm cap}$, propulsion at velocities
near $V_{\rm obst}$ is possible, and in this case indirect
processes dominate. Using very large values
of $k_{\rm br}$ will also yield an asymptotic velocity of
$V_{\rm max}$, since the relevant quantity in balancing
branching with capping 
is the ratio of the branching to the capping rates. In this
case there will be an added effect from the decrease of the
average branch spacing, which will also increase the fraction
of indirect branching events.}
\end{itemize}
%

The above argument depends on the ratio of $d$ to
the relative branch spacing. As $d$ becomes smaller, the magnitude
of the indirect branching terms becomes less. One should then
have to go to progressively smaller values of $k_{\rm cap}$ to 
reach the regime where indirect branching dominates
and the velocity approaches $V_{\rm max}$. This is
demonstrated in Figure 5, which shows $V_{\rm obst}$ vs. 
$k_{\rm cap}$ for a range of values of $d$ ranging up 
to $10a$, in the three-dimensional side-branching model. 
The branching layer thickness cannot be much greater than
$10a$, because this would lead to exponential growth
in the filament density away from the obstacle, and this
has not been observed. It is seen in Figure 5 that the
growth velocity varies in a fairly linear fashion with $k_{\rm cap}$
down to a crossover value $k_{\rm cap}^{\rm c}$,
at which it turns upward; $k_{\rm cap}^{\rm c}$
increases with $d$. $k_{\rm cap}^{\rm c}$ should be 
proportional to the rate of
indirect branching events. The latter is proportional to both the
rate of production $k_{\rm br}$ of new filaments and the 
fraction of these new filaments which branch before they 
leave the branching region. Since most of the new filaments
will point at angles relatively far from the growth direction, 
they will leave the branching region rapidly. The fraction
that branch before this happens will be proportional to
$d/l_{\rm br}$, where $l_{\rm br}$ is the average spacing
between branches along a filament. Thus
\begin{equation}
k_{\rm cap}^{\rm c} = \alpha k_{\rm br} d/l_{\rm br}
\label{kcapc}
\end{equation}
where, $\alpha$ is a dimensionless constant. From our numerical
results, we find that $\alpha=0.2$ and $\alpha=0.4$ for side
and end branching, respectively, in three dimensions.
The possibility of applying this effect experimentally is
evaluated in the {\bf Discussion} section.

\section{Nucleation Model}

In nucleation models, the obstacle generates new filaments 
without making use of the existing filament network, and it is
assumed that the generated filaments subsequently attach to
to this framework.  We thus take the overall generation rate 
for new filaments to be independent of the filament density.
However, it is not possible for the orientation distribution 
of new branches to be independent of $n(\theta ,t)$, since the
new branches must satisfy the $70^{\circ}$ branching angle 
constraint. For this reason, we obtain the equation of motion
for the nucleation model by dividing the first term in
Eq. (\ref{dndt}) by the total rate of new filament generation,
so that
\begin{eqnarray}
 {\partial n(\theta,t) \over \partial t} &=& 
{k_{\rm nuc} \over B} 
\int_{0}^{\theta_{\rm max}}
D(\theta ,\theta ') \nu(\theta') n(\theta ',t) d\theta '
-k_{\rm cap}n(\theta ,t)
\nonumber  \\
 & & - H[V_{\rm obst}-v(\theta)]
[(V_{\rm obst}-v(\theta))/d]n(\theta,t),
\label{dndtnuc}
\end{eqnarray}
where $k_{\rm nuc}$ is total the number of filaments generated
per unit time, 

\noindent{$B= \int_{0}^{\theta_{\rm max}}
D(\theta ,\theta ') \nu(\theta') n(\theta ',t) d\theta d\theta '
\times ({\rm area ~ or ~ length})$,
and the other quantitites and parameters
are as in the autocatalytic model. The area factor is used for
three dimensions, the length factor in two dimensions. We ignore
the correction used in Eq. (\ref{dndt}) to account for exponential 
growth of the filament density, because in nucleation models 
this does not occur.}

This rate equation is solved by numerically stepping forward
in time, at a fixed obstacle velocity, until a steady-state
filament orientation distribution $n(\theta)$ is obtained. 
The force is then obtained via Eq. (\ref{ftot}). 
In order to evaluate the right-hand side of Eq. (\ref{ftot}),
it is necessary to take a specific form for
the force-velocity relation of a single filament. We take the form
suggested by Brownian-ratchet theory\cite{Peskin93,Mogilner96}:
\begin{eqnarray}
f(v,V_{\rm obst}) = & \frac{kT}{a} {V_{\rm max} \over v} 
ln(v/V_{\rm obst}) 
& (v \ge V_{\rm obst}) \label{fofv} \\
f(v,V_{\rm obst}) = & 0 & (v < V_{\rm obst}),
\label{fofv1}
\end{eqnarray}
which translates to an exponential dependence when the
velocity is given in terms of force. Here the factor of
${V_{\rm max} \over v} = {1 \over \cos{\theta}}$ 
accounts for the orientation dependence
of the step size per monomer.

Figure 6 shows the calculated force-velocity relation for
the network in three-dimensional side-and end-branching 
nucleation models. 
The parameters,
$k_{\rm cap} = 0.35 sec^{-1}$ and $k_{\rm br} = V_{\rm max}/20a$, 
are the same as in Figures 3 and 4. To evaluate $k_{\rm nuc}$
we use the value $k_{\rm nuc}/k_{\rm cap} = 100$ suggested by
experiments\cite{Kuo00} on {\it Listeria}; 
similar results are obtained for the value
$k_{\rm nuc}/k_{\rm cap} = 10$ suggested by experiments\cite{Cameron99} on
beads, except that the horizontal scale is compressed. 
For comparison, we include the force-independent behavior
found in the autocatalytic models.
The network force-velocity relation in the nucleation models 
differs from the exponential decay
for a single filament in two ways: 
\begin{itemize}
\item{There is a very rapid dropoff in the low-force
region, where $V_{\rm obst}$ is near $V_{\rm max}$. For 
$V_{\rm obst} \simeq V_{\rm max}$, 
the number of filaments is reduced by leaving effects, and 
only the fraction of filaments with near-optimal orientations contact
the obstacle. Thus only a very small number of filaments contact the
obstacle. This leads to a rapid decrease of the velocity
with applied force, since the force per filament is large.}
\item{There is a decay at large forces, but it 
is slower than exponential. 
This occurs because with increasing force the load is redistributed 
between filaments of different orientations. 
For small forces, filaments with $\theta \simeq 0$ carry
most of the load, because they are the only ones in contact
with the obstacle. However, for larger forces, 
the most rapidly growing filaments are those with larger values
of $\theta$, as was pointed out in 
previous work developing the Brownian-ratchet
model for single filaments\cite{Mogilner96}.  
The velocities of these filaments decay less rapidly with
force, because their orientation gives a smaller step size per monomer. 
At large forces, filaments with progressively smaller values of
$v$ dominate, reducing the decay rate of the velocity with applied
force.} 
\end{itemize}
The overall shape of the force-velocity relation is quite similar 
to that obtained by a ``tethered-ratchet" model\cite{Mogilner03a} 
which treats two types of filaments, ``attached" and ``working", 
where the working filaments supply the motile force by 
polymerization. In this model the deviations from exponential 
behavior result from changes in the relative numbers of 
attached and working filaments, an effect not included in the 
present model. 

We do not show detailed results for the dependence of
the velocity in nucleation models
on the rate parameters $k_{\rm br}$ and $k_{\rm cap}$. 
However, the main findings are that the dropoff of velocity
with increasing $k_{\rm cap}$ is steeper than in autocatalytic
models, and its dropoff with decreasing $k_{\rm nuc}$ is
more rapid than its dropoff with decreasing $k_{\rm br}$ in
autocatalytic models.


\section{Sensitivity to Key Approximations}

The preceding sections have presented calculated growth 
velocities for several models for branch generation
during actin-based motility, which differ
in their underyling assumptions and give distinct
results. We now discuss how these results depend on the 
approximations and assumptions made in the models. 
The most important of these are the following:  

\begin{itemize}
\item{Neglect of filament-filament
interactions. The stochastic simulations (\onlinecite{Carlsson01}) 
showed that at 
typical polymerized-actin densities, steric volume exclusion
has only a small impact. Electrostatic interactions are also 
expected to have minor effects because the Debye screening length 
of 1 nm at typical physiological ionic strengths of 
150 mM is much less than the typical filament spacing\cite{Abraham99} 
of 30nm. Thus the neglect of filament-filament interactions 
in the above models seems to be a reasonable approximation.
Inclusion of these interactions would cause the velocity to
be reduced at large branching rates, small capping rates, and
high forces in autocatalytic models. The effects in
nucleation models would be weaker because the density of filaments 
approaches a constant value at high forces.}
\item{Neglect of filament bending and branch-point elasticity. These
effects could lead to individual filaments changing their orientation
over time. However, as discussed above, typical branch-point angle fluctuations
are about $10^{\circ}$.  Provided that filaments remain short, the angle
fluctuations from filament bending are roughly the same as
those from the branch points\cite{Carlsson01}. Thus the changes in filament
orientation should not have a major impact on the results.
Bending and elasticity could also have a substantial impact on
the single filament force-velocity relation. However, previous work\cite{Mogilner96}
has argued that the exponential form continues to hold when filament elasticity
is included. 

The actin network can also propagate effective elastic interactions from
one point to another. These could lead to long-ranged effective interactions between
filaments. The major effect of such interactions would be a stiffening which would
increase with increasing filament density. Such a stiffening would reduce the
filament-end fluctuations in a density-dependent fashion. 
This would lead to a reduction in network growth 
velocity with increasing density. However, this effect is expected to be 
small, because at observed filament lengths, the thermal fluctuations 
of individual filaments are already equal to or greater than the monomer size;
the network elasticity would only serve to enhance these.}
\item{Neglect of severing and annealing effects. These have 
been treated previously
in a rate-equation model\cite{Sept99}. The rates obtained there
are much lower than the capping and branching rates used here. 
However, if there is a very large acceleration of severing and/or
nucleation in the cellular environment, these effects could become
important. 
If filament severing is independent of interactions between filaments, 
the severing terms are linear in the filament concentration. Thus they 
would add a linear 
term to Eq. (\ref{dndt}). Such a term would transfer free ends in the network 
from the branching region to regions farther from the obstacle, since the leftover 
filament free barbed ends would generally be outside the branching region. 
The severing effects would thus in some ways act like an increase in 
$k_{\rm cap}$ or the leaving terms. 
The force-independence of the velocity in the
autocatalytic models would continue to hold because it depends only on the linearity
of the rate equation. However, in the nucleation models, inclusion of severing
would have the effect of accelerating the dropoff of velocity with applied force. 

On the other hand, annealing corresponds to nonlinear terms\cite{Sept99}
in the rate equation. The most important type of event would be the 
incorporation of filament fragments into the network. The resulting effects 
on the branching region would be small, because the fragments would be
overwhelmingly capped. Even the uncapped ones would have a small
effect because the likelihood of their free ends being inside the branching
region after network incorporation would be small. In addition, the time
scales for annealing were found\cite{Sept99} to be on the order of several
hours, and thus they are likely too slow to be important here.}
\item{Filament uncapping and branch detachment. Spontaneous uncapping
rates are estimated\cite{Schafer96} to be in the range of $10^{-4}s^{-1}$.
Thus very little uncapping would occur during the time that a capped filament
spends in the branching region. However, obstacle-induced filament
uncapping could occur more rapidly, and this would change the 
growth velocities. Branch detachment rates are related to the
decay of the filamentous-actin density away from the obstacle. 
Observations of {\it Listeria} tails\cite{Tilney94} and 
tails on beads mimicking {\it Listeria}\cite{Cameron01} 
indicate that the
tail density decays over a distance of microns away from the obstacle,
and similar results are obtained for the network density around
lamellipodia\cite{Svitkina99}. Thus little branch detachment is expected 
over the thickness of the branching region.}
\item{Restrictions on the orientation of new filaments. Our earlier
simulations\cite{Carlsson01} had suggested that observed filament
structures near the growth front are better described if new branches
are allowed only in the forward direction. We have performed
runs including such effects, and find curves quite similar to
those shown above. Provided that $k_{\rm br}$ is 
adjusted to keep the filament generation rate constant,
the main effect of the orientation restriction is a moderate
increase in the growth velocity.}
\item{ATP hydrolysis. Hydrolysis of a filament monomer generally is believed 
to occur on a time scale of several seconds\cite{Blanchoin02}, and by this time 
the branching region will have moved away from the monomer.}
\item{Depletion of actin and actin-binding proteins. These effects would change
the rate parameters in Eq. (\ref{dndt}), as well as $V_{\rm max}$
and the force-velocity relation. A previous analysis of
this issue\cite{Carlsson01} showed
that for obstacles of size up to $1 \mu m$, the depletion effects are 
less than $20\%$. For larger obstacles, the effects
can be more significant.}
\item{Rate-limiting activation steps upstream of the branch-generation
step. If the Arp2/3 activation process has a long activation time, then
there will be a limit to the number of filaments that can be generated
per unit time per unit area of the obstacle. The presence of such
activation steps would result in a behavior similar to that of nucleation
models, even if preexisting branches are required for new branch nucleation.}
\item{Effects of fluctuations due to small numbers of filaments. We expect
these to be proportional to $\sqrt{N_{\rm tot}}$. Taking 20\% as a cutoff for
fluctuations, substantial corrections to the present results
would begin to set in at $N_{\rm tot}=25$.}
\item{The relation (\ref{fofv})  between force and filament velocity, 
used in the nucleation model.
This relation assumes an exponential dependence of the velocity on
applied force, and a particular exponential decay parameter. While such
a relationship has been found in model calculations\cite{Peskin93,Mogilner96}
and Brownian-dynamics simulations\cite{Carlsson00}, the true relation may be 
more complex or have a different decay parameter. In addition, Eq. (\ref{fofv})
ignores potential attachment forces between the filaments and the obstacle. 
The presence of such attachments in the case of 
{\it Listeria} has been demonstrated by attempts to detach 
the bacterium from its tail using optical tweezers\cite{Gerbal00}, and by
measurements of the bacterium position\cite{Kuo00}
which have found very small fluctuations. In the case of beads,
the presence of attachments is demonstrated by the continuous motion of 
a 50 nm bead propelled by a single filament\cite{Cameron01}; 
without attachments, the bead would rapidly diffuse 
away from the filament tip. The results for the
velocity in the autocatalytic model would continue to hold 
regardless of the attachment force, since the force does
not enter these calculations. The results
for the nucleation model would be strongly influenced by 
attachments, although the nature and magnitude of the effects
are not certain. As mentioned above,
a recent model\cite{Mogilner03a} has treated 
the effects of filament attachments on
the force-velocity relation for {\it Listeria}, and found that
the attachments tend to accentuate the dropoff at small forces,
and reduce the dropoff at high velocities. Thus they tend to reinforce
the behavior found here, and could amplify the differences between
the autocatalytic and nucleation models. }
\end{itemize}

\section{Discussion}

The above analysis has shown that both the dependence of 
the growth velocity on key protein concentrations and
the force-velocity relation are sensitive to the details
of the generation process for new branches. This motivates
measurements of these dependences.
Measurement of the dependence of the growth velocity on
the activated Arp2/3 and capping-protein (CP) concentrations would
require the use of a pure-protein medium in order to control
secondary effects from the concentrations of other proteins. 
Such media have been used in studies of both 
{\it Listeria}\cite{Loisel99} and plastic beads coated with 
VCA\cite{Bernheim02}. In pure-protein media, one could monitor
the free-actin concentration in the growth medium as the
concentrations of activated Arp2/3 and capping protein are
changed, and buffer the actin appropriately to keep the free-actin
concentration constant. Measurements of the dependence of the
growth velocity on the Arp2/3 concentration at fixed CP
concentration could shed light on the relative importance
of side and end branching. As indicated in Figure 3, end-branching
models would lead to a sharp cutoff in growth when the 
Arp2/3 concentration drops below a critical value, while
side-branching models would display a much more gradual 
cutoff. Measurements of the dependence of the growth velocity
on CP concentration at fixed Arp2/3 concentration could,
in principle, help
establish the thickness of the branching region. The results in
Figure 5 show that for a branching region less than a single
monomer in thickness, the plot is essentially a straight line 
as the CP concentration goes to zero. If the branching layer
thickness is one or a few monomers, the velocity displays a sharp
upturn at small CP concentrations. If the thickness is greater than
a few monomers, the velocity curves smoothly upwards as the
CP concentration drops. However, because the magnitude
of the differences between the curves is fairly small, obtaining
velocity measurements of sufficient resolution to assess the
branching layer thickness might be impossible.

Quantitative measurement of the force-velocity
relation would appear to be the most straightforward way of
using the present results to evaluate competing models of
filament generation.  Such experiments have been performed
by using methylcellulose to vary the
viscosity, for beads moving 
in pure-protein media\cite{Wiesner03} and 
bacteria moving in cell extracts\cite{McGrath03}. 
Because of the simplicity of the growth medium and moving
obstacle, the bead experiments would appear to be the closest
to the present calculations. These experiments indicated that, over a broad
range of forces up to approximately 50 pN, the velocities
of 2$\mu$m beads varied by only 30\%. This behavior is  
consistent with the autocatalytic model discussed above. However,
we cannot yet draw a definite conclusion because the nucleation
rate in the nucleation model is not firmly established, and with
a very high nucleation rate the velocity could be insensitive to
force up to 50 pN. The measurements of bacterial motion found
that the velocity at 50 pN opposing force is much less than its
value at zero opposing force. These results would suggest that
the generation rate of new filaments is limited,
perhaps because of the presence of different rate-limiting steps
than for the beads. We note, however, that in the bead experiments,
a correction for the effects of methylcellulose not related to
viscosity was made, and no such correction was made in the 
bacterium experiments. This could also be an important factor
in explaining the differences between the results.

Other possible methods for measuring the force-velocity relation 
involve the use of laser-based optical-tweezer techniques.
In such methods
one tracks a fluorescently labeled object (bead or bacterium), 
and the force is determined by the position of the object relative
to the center of the laser spot. One can then impose a feedback
loop which keeps the force fixed by motion of the substrate, 
and then measure the velocity
by tracking the object's coordinates. This method would avoid 
any uncertainties resulting from the addition of thickeners to
the cell extract. However, because optical tweezers are only able
to exert forces up to about 40 pN, it would be necessary to use 
conditions under which not too many filaments impinge on the object. 

Two other types of experiments in the literature have some relevance 
to our results. The first involves attempts to stop the motion of
{\it Listeria} with an optical trap\cite{Gerbal00}.
These experiments found that the trap could temporarily stop
the motion, but the bacterium eventually broke free due
to an increase in the force supplied by the tail. However,
because the force exerted by the trap is only about
10 pN, these experiments are unable to distinguish between
the models considered here. The second
treats the dynamics of the actin filament density behind ``hopping"
{\it Listeria}. These are mutants in which roughly 80 residues of the
ActA surface protein have been deleted. They move rapidly for short
intervals of time, stop for longer periods, then  move again and
repeat the cycle. Experiments with fluorescent 
actin\cite{Lasa97,Fung98}
have shown that the fluorescence intensity increases during the stationary
period, suggesting that the actin density is building up to counter
the forces opposing the motion of the bacterium. However,
the opposing force is not known in these experiments.

\section{Acknowledgements}

I am grateful to John Cooper, David Sept, and Jonathan
Katz for informative 
conversations.  This research was supported by the National 
Institutes of Health under Grant Number GM38542.

\newpage
\begin{table}
\centerline{Figure Captions}
\bigskip
\bigskip
\noindent 
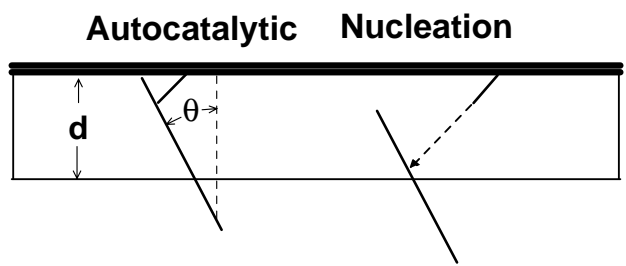
Figure 1: Schematic of autocatalytic vs. nucleation-based
branch-generation processes. $d$: branching layer thickness.
$\theta$: angle between filament and growth direction.

\bigskip

\noindent 
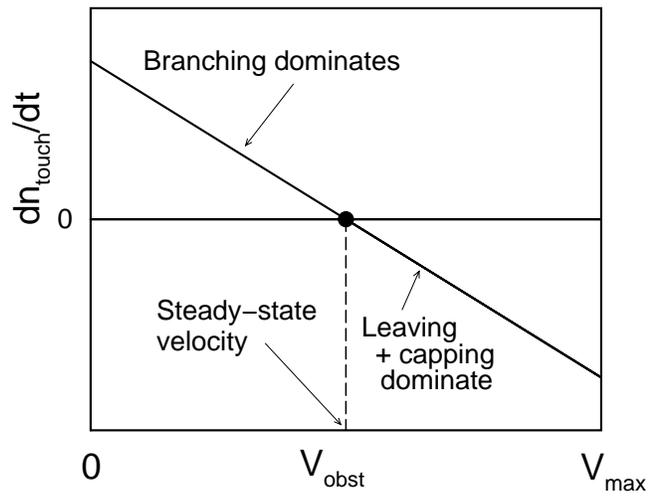
Figure 2: Mechanism determining steady-state obstacle
velocity $V_{\rm obst}$ in autocatalytic models. 
$V_{\rm max}$: free-filament growth velocity.
$n_{\rm touch}$: total number of filaments in branching region.
Steady-state velocity  is that for which 
$dn_{\rm touch}/dt$ = 0.

\bigskip

\noindent 
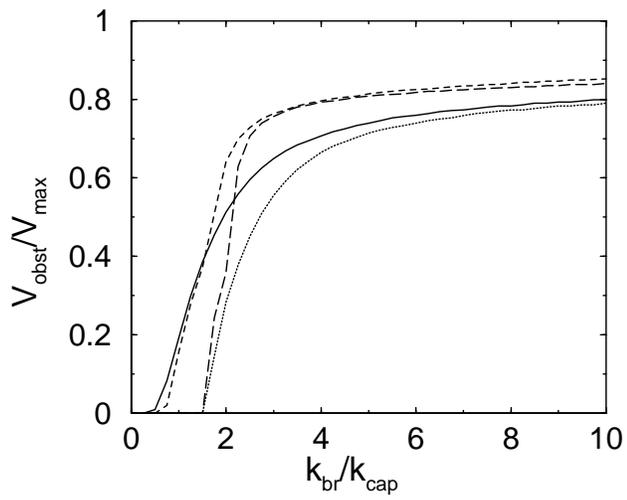
Figure 3: Dependence of steady-state obstacle velocity 
$V_{\rm obst}$ on branching rate $k_{\rm br}$, with
$k_{\rm cap}$ fixed at 0.35 $sec^{-1}$.
$V_{\rm max}$: maximum projected filament velocity.
$k_{\rm cap}$: capping rate.
Solid line: side-branching model in three dimensions. Dotted line:
end-branching model in three dimensions. Dashed line: 
side-branching model in two dimensions. Long-dashed line: end-branching
model in two dimensions.

\bigskip

\noindent 
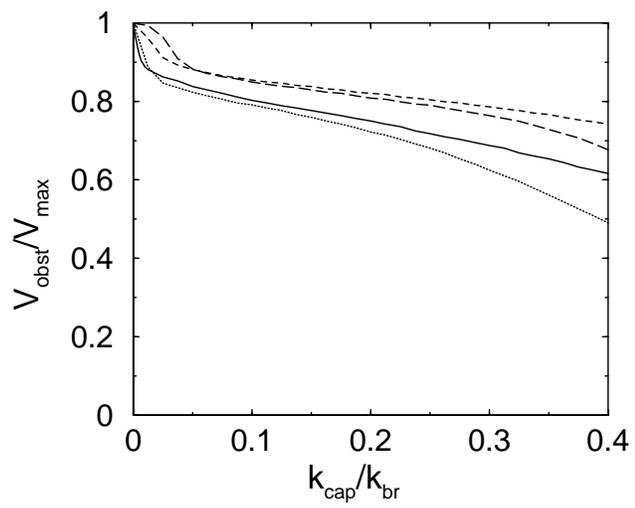
Figure 4: Dependence of steady-state obstacle velocity 
$V_{\rm obst}$ on capping rate $k_{\rm cap}$, 
with $k_{\rm br}$ fixed at 
$V_{\rm max}/20a$=5 sec$^{-1}$.
$V_{\rm max}$: 
maximum projected filament velocity.
$k_{\rm br}$: branching rate.
Solid line: side-branching model in three dimensions. 
Dotted line: end-branching model in three dimensions. 
Dashed line: side-branching model in two dimensions. 
Long-dashed line: end-branching model in two dimensions. 

\bigskip 

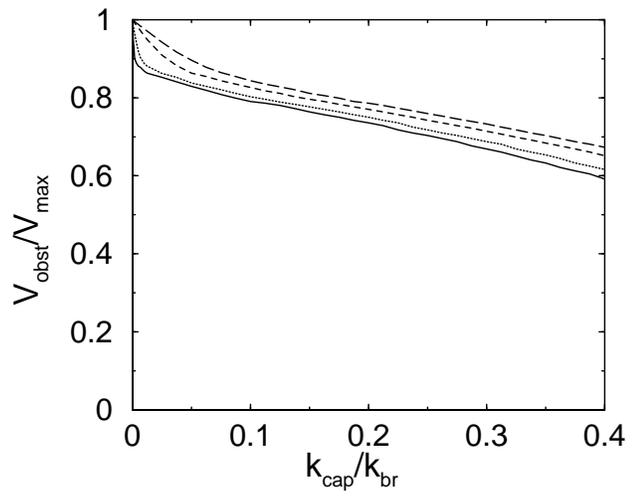
\noindent Figure 5: Effect of branching layer thickness $d$ 
on $k_{\rm cap}$-dependence of obstacle velocity
$V_{\rm obst}$,  
with $k_{\rm br}$ fixed at 
$V_{\rm max}/20a$=5 sec$^{-1}$.
$V_{\rm max}$: 
maximum projected filament velocity.
$k_{\rm br}$: branching rate.
Solid line: $d$=0.1a (a=monomer size). Dotted line:
$d$=a. Dashed line: $d$=5a. 
Long-dashed line: $d$=10a. 

\bigskip

\noindent 
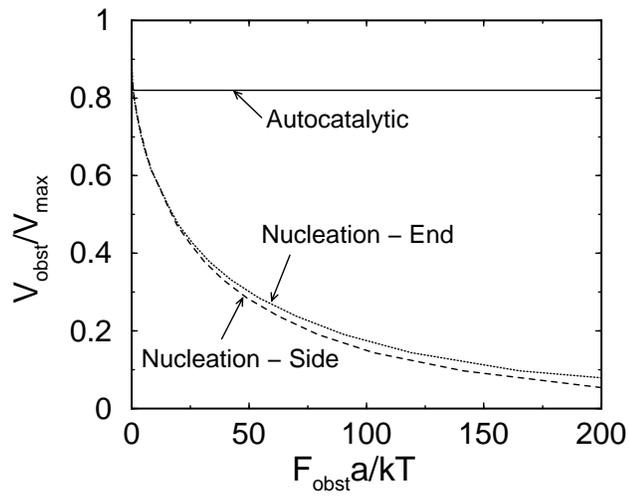
Figure 6: Force-velocity relation for nucleation and autocatalytic
models, with side and end branching. 
Rate parameters are
$k_{\rm cap}$ = 0.35 $sec^{-1}$ and
$k_{\rm br}$ = 
$V_{\rm max}/20a$=5 sec$^{-1}$.
$V_{\rm obst}$: obstacle velocity. $V_{\rm max}$: maximum
free-filament velocity. $F_{\rm obst}$: force exerted by 
filaments on obstacle. $a$: step size along filament. 
\end{table}
\end{document}